\documentclass[prd,twocolumn]{revtex4}\begin{document}\preprint{}
\title{ Stochastic wave equation with thermal noise in an expanding universe }

\author{Z. Haba \\
Institute of Theoretical Physics, University of Wroclaw,\\ 50-204
Wroclaw, Plac Maxa Borna 9, Poland}
\email{zhab@ift.uni.wroc.pl}\date{\today}

\begin{abstract} We discuss Einstein-Klein-Gordon system in an
environment of an infinite number of scalar fields leading to an
external thermal noise. In the lowest order of    metric and field
perturbations the quadratic fluctuations consist of a sum of
quantum and thermal fluctuations. We show that these fluctuations
depend on the form of the interaction of the inflaton with the
environment.
\end{abstract}
\maketitle
  \section{Introduction}
We consider a model of Einstein gravity with an infinite number of
scalar fields (in addition to the inflaton which generates the
inflation). Such a model has been discussed in
\cite{berera}\cite{adv}. In a Markovian approximation the dynamics
of the infinite number of fields with some unknown initial
(random) values can be  approximated by a random force according
to the well-known scheme of Brownian motion \cite{kac}. In
comparison to the standard model of inflation (cold inflation) an
extra thermal noise appears (accompanied with a friction term) on
the rhs of the wave equation as an expression of the influence of
an infinite number of unobserved scalar fields. The unobserved
scalar fields on the rhs of Einstein equations are also replaced
by a noise term in such a way that the conservation law of the
total energy-momentum (required for the consistency of Einstein
equations) is satisfied. As in the standard perturbation expansion
\cite{lifszyc}\cite{mukhanov}\cite{sakai}
\cite{basset}\cite{marozzi}\cite{hwang} we can express the
perturbations of the metric appearing in the wave equation of the
inflas of the scalar fields. In this way we
obtain a linearized stochastic wave equation for inflaton
fluctuations dependent on the thermal noise and on the solutions
of the time-dependent Einstein-Klein-Gordon equations. The
solution of the wave equation is a superposition of a solution of
the homogeneous equation (without source) and a solution of the
inhomogeneous equation with the thermal noise as a source. We
quantize the solution of the  homogeneous equation. Then, the
quadratic fluctuations consist of quantum fluctuations and thermal
fluctuations induced by the thermal noise. We calculate these
fluctuations and explore their dependence on the interaction with
the environment.

\section{Einstein equations}
The  energy-momentum tensor of the inflaton $T_{\mu\nu}$ in the
presence of other scalar fields is not conserved. We have to
compensate  the energy-momentum by means of a compensating
energy-momentum $T_{de}$ which we associate with the dark sector
so that $T_{tot}^{\mu\nu}$
\begin{equation} T_{tot}^{\mu\nu}=T^{\mu\nu}+T^{\mu\nu}_{de}\end{equation}
 is conserved .

 The energy-momentum  is assumed to be in the form of
 an ideal fluid
\begin{equation}
T^{\mu\nu}_{de}=(\rho_{de}+p_{de})u^{\mu}u^{\nu}-g^{\mu\nu}p_{de},
\end{equation}where $\rho$ is the energy density and $p$ is the
pressure.  The velocity $u^{\mu}$ satisfies the normalization
condition
\begin{displaymath}  g_{\mu\nu}u^{\mu}u^{\nu}=1.\end{displaymath}

In the case of the inflaton we have the representation
\begin{equation}
u^{\mu}=\partial^{\mu}\phi(\partial^{\sigma}\phi\partial_{\sigma}\phi)^{-\frac{1}{2}},
\end{equation}
\begin{equation}
\rho+p=\partial^{\sigma}\phi\partial_{\sigma}\phi,
\end{equation}
\begin{equation}p=\frac{1}{2}\partial^{\sigma}\phi\partial_{\sigma}\phi-V.
\end{equation}
 We consider flat FLWR metric
 \begin{equation}
 ds^{2}=dt^{2}-a^{2}d{\bf x}^{2}
 \end{equation}Einstein
equations are written in the form
\begin{equation}
 R^{\mu\nu}-\frac{1}{2}g^{\mu\nu}R=8\pi G
T_{tot}^{\mu\nu},
\end{equation} where $G$ is the Newton constant.

The Friedman equation in the FRLW flat metric  has the form
\begin{equation}\begin{array}{l}
H^{2}=\frac{8\pi G}{3}(\rho+\rho_{de}).\end{array}\end{equation}

\section{Expansion around the homogeneous solution}
In \cite{berera}\cite{adv} a linear interaction with an infinite
set of environmental scalar fields with a coupling $\lambda_{a}$
and frequencies $\omega_{a} $ has been studied. If
$\lambda_{a}=\gamma \pi^{-\frac{1}{2}}\omega_{a}$ (there is an
error in eq.(16) of ref.\cite{adv} in this relation) then the
inflaton equation acquires  a friction $\gamma$ and a mass-like
term $\gamma^{2}\frac{3}{2}H\phi$. Then, on a flat FLWR metric (6)
the inflaton  equation reads
\begin{equation}\begin{array}{l}
\partial_{t}^{2}\phi_{c}-a^{-2}\triangle\phi_{c}+(3H+\gamma^{2})\partial_{t}\phi_{c}
+V^{\prime}(\phi_{c})+\frac{3}{2}\gamma^{2}H\phi_{c}=0.
 \end{array}\end{equation}
 If the initial values of the environmental fields have the Gibbs distribution
 with a temperature $\beta^{-1}$ then the random fields sum up  to a noise. Then, eq.(9) with the noise (satisfying the fluctuation-dissipation
 theorem) on a manifold with the metric $g_{\mu\nu}$, according to the derivation in \cite{berera}\cite{adv}, takes the form
\begin{equation}\begin{array}{l}
g^{-\frac{1}{2}}\partial_{\mu}g^{\frac{1}{2}}\partial^{\mu}\phi+\gamma^{2}\partial_{t}\phi
+V^{\prime}(\phi)+\sigma\frac{3}{2}\gamma^{2}H\phi=\beta^{-\frac{1}{2}}\gamma
a^{-\frac{3}{2}}W,\end{array}
 \end{equation}where $g=\vert \det(g_{\mu\nu})\vert$ and
 $\sigma=1$. Eq.(10) with $\sigma=0$ ( $\gamma$ for general interactions with the environment may depend on $\phi$ ) is treated as
a model of warm inflation \cite{warm}. It describes the universe
evolution if thermalization occurred before inflation. We neglect
the dependence of $\gamma$ on $\phi$ but take into account the
possible appearance of the mass-like correction
$\sigma\frac{3}{2}\gamma^{2}H\phi$ from the interaction with the
environment (as in models of \cite{berera}\cite{adv}).

 The white noise $W$ in eq.(10) is the Gaussian process (related to the Brownian motion $B$) with the covariance
 \begin{equation}
 \langle W_{t}({\bf x})W_{s}({\bf y})\rangle dt=  \langle dB_{t}({\bf x})dB_{s}({\bf y})\rangle
 =\delta(t-s)\delta({\bf x}-{\bf y})dt
\end{equation}
From the definitions (2)-(5) of the inflaton energy-momentum and
eq.(10) we obtain the (non)conservation law
 \begin{equation}
 (T^{\mu\nu})_{;\mu}=\partial^{\nu}\phi(\gamma
\beta^{-\frac{1}{2}}
a^{-\frac{3}{2}}W-\gamma^{2}\partial_{t}\phi-\sigma\frac{3}{2}\gamma^{2}H\phi)
 \end{equation}
The zero component part  $T^{0\nu}$ of eq.(12) is interpreted as a
stochastic differential equation in the sense of Stratonovich
\cite{ikeda}(the circle denotes the Stratonovitch multiplication
of the Brownian differentials $dB$)

 \begin{equation}\begin{array}{l} d\rho+3(1+w_{I})H\rho dt=\gamma\beta^{-\frac{1}{2}}
\partial_{t}\phi\circ a^{-\frac{3}{2}} dB
\cr-\gamma^{2}(\partial_{t}\phi)^{2}dt-\sigma\frac{3}{2}\gamma^{2}H\phi\partial_{t}\phi
dt, \end{array}\end{equation}
 where
\begin{equation} w_{I}=
(\frac{1}{2}(\partial_{t}\phi)^{2}-V)(\frac{1}{2}(\partial_{t}\phi)^{2}+V)^{-1}.
 \end{equation}  From the conservation law

\begin{equation}
(T_{de}^{\mu\nu})_{;\mu}=-(T^{\mu\nu})_{;\mu}\end{equation}
 the compensating energy density must have the (non)conservation
law with an opposite sign
\begin{equation}\begin{array}{l}
d\rho_{de}+3H(1+w)\rho_{de}dt=-\gamma\partial_{t}\phi
 \beta^{-\frac{1}{2}}a^{-\frac{3}{2}}\circ
dB\cr+\gamma^{2}(\partial_{t}\phi)^{2}dt+\sigma\frac{3}{2}\gamma^{2}H\phi\partial_{t}\phi
dt\cr,\end{array}
\end{equation}where
\begin{equation}
w=\frac{p_{de}}{\rho_{de}}.\end{equation} Eqs.(16)-(17) determine
$\rho_{de}$ and the energy-momentum tensor of the ideal fluid (2).
We write
  \begin{equation}
  \phi=\phi_{c}+\delta\phi
  \end{equation}
We perturb the metric $g_{\mu\nu}$ around the flat FLWR metric (6)
in the uniform curvature gauge \cite{marozzi}\cite{hwang} and
eliminate the perturbed metric from the Einstein-Klein-Gordon
system (7) and (10) expressing the metric perturbations by
perturbations of the scalar fields. Then, in the linear
approximation for the inflaton perturbation $\delta\phi$ we get
the equation (at $\gamma=0$ this is eq.(102) of \cite{basset})

\begin{equation}\begin{array}{l}
\partial_{t}^{2}\delta\phi-a^{-2}\triangle\delta\phi+(3H+\gamma^{2})\partial_{t}\delta
\phi\cr+V^{\prime\prime}(\phi_{c})\delta\phi-6\epsilon
H^{2}\delta\phi+\sigma\frac{3}{2}\gamma^{2}H\delta\phi=\gamma
\beta^{-\frac{1}{2}} a^{-\frac{3}{2}}W_{t},\end{array}
 \end{equation}where
 \begin{equation}
 \epsilon=-H^{-2}\partial_{t}H.
 \end{equation}

\section{Power spectrum of the linearized wave equation}
We introduce the conformal time
\begin{equation}
\tau=\int dt a^{-1}.
\end{equation}With a slowly varying $H$ we have approximately
\begin{equation}
aH=-(1-\epsilon)^{-1}\frac{1}{\tau}.
\end{equation}
as a consequence of an integration of the identity \cite{wood}
\begin{displaymath}
\partial_{t}\Big((1-\epsilon)Ha\Big)^{-1}=-a^{-1}+\partial_{t}\epsilon
\Big(aH(1-\epsilon)^{2}\Big)^{-1}
\end{displaymath}

 In terms of $\tau$
eq.(19) for the Fourier transform $\delta\phi({\bf k})$ reads
($k=\vert{\bf k}\vert$)
\begin{equation}\begin{array}{l}
(\partial_{\tau}^{2}-\frac{2+3\Gamma}{1-\epsilon}\frac{1}{\tau}\partial_{\tau}+k^{2}
+\frac{3\eta-6\epsilon+\frac{9}{2}\sigma\Gamma}{(1-\epsilon)^{2}}\tau^{-2})\delta\phi=\gamma\beta^{-\frac{1}{2}}
W_{\tau},\end{array}
 \end{equation}
where $\sigma$ is equal $1$ in the model \cite{berera} \cite{adv}
and $\sigma=0$  in some models of warm inflation
\cite{warm}\begin{equation} 3\eta=V^{\prime\prime}H^{-2},
\end{equation}and
\begin{equation}
\Gamma=\frac{\gamma^{2}}{3H}.
\end{equation}
In eq.(23) we applied  the transformation  property of the white
noise $\sqrt{a}W_{t}=W_{\tau}$.

Let
\begin{equation}
\delta\phi=\tau^{\alpha}\Psi
\end{equation}
with
\begin{equation}
\alpha=\frac{1+\frac{3}{2}\Gamma}{1-\epsilon}.
\end{equation}
Then
\begin{equation}\begin{array}{l}
(\partial_{\tau}^{2}+k^{2}
+\frac{-2+3\eta-5\epsilon-\frac{9}{4}\Gamma(\Gamma+\frac{2}{3}-2\sigma)
+\frac{3}{2}\epsilon\Gamma}{(1-\epsilon)^{2}}\tau^{-2})\Psi\cr=\gamma
\beta^{-\frac{1}{2}} \tau^{-\alpha}W_{\tau}.\end{array}
 \end{equation}
 The lhs of this equation agrees with Bassett et al \cite{basset} (eq.(103)) for $\gamma=0$.
 Let us still use another form of the stochastic equation.
 Let
 \begin{equation}
 \delta\phi=\tau^{\mu}\Phi
 \end{equation}
 with
\begin{equation}
\mu=(1-\epsilon)^{-1}(\frac{3}{2}-\frac{\epsilon}{2}+\frac{3}{2}\Gamma)
\end{equation}
Then
\begin{equation}
(\partial_{\tau}^{2}+\tau^{-1}\partial_{\tau}+(k^{2}-\nu^{2}\tau^{-2}))\Phi=\gamma
\beta^{-\frac{1}{2}} \tau^{-\mu}W_{\tau}
\end{equation}
where
\begin{equation}
\nu^{2}=(1-\epsilon)^{-2}\Big(\frac{9}{4}-3\eta+\frac{9}{2}\epsilon+\frac{9\Gamma(1-\sigma)}{2}
+\frac{1}{4}(\epsilon-3\Gamma)^{2}\Big)
\end{equation}
Without noise the solution of eq.(31) is the Hankel function
$H_{\nu}(k\tau)$. The solution of eq.(28) without noise is
$\psi_{\nu}=\tau^{-\alpha+\beta}H_{\nu}$. Then, the solution of
the stochastic wave equation for $\Psi$ is

\begin{equation}\begin{array}{l}
\Psi(\zeta)=\gamma k^{-2}\int_{\zeta}^{\infty} {\cal
G}(\zeta,\zeta^{\prime})\zeta^{\prime-\alpha}k^{\alpha}\sqrt{k}W_{\zeta^{\prime}}d\zeta^{\prime}
\end{array}\end{equation} where ${\cal G}$ is the Green function for the operator on the lhs of eq.(28) and
\begin{equation}
\zeta=k\tau
\end{equation}
The Green function can be constructed \cite{roach} from the two
independent solutions of the homogeneous equation (28)
 (without noise)
\begin{equation}
\psi_{1}=\zeta^{-\alpha+\mu}J_{\nu}(\zeta)
\end{equation}\begin{equation}
\psi_{2}=\zeta^{-\alpha+\mu}Y_{\nu}(\zeta)
\end{equation} where the Bessel functions $J$ and $Y$ can be defined by the Hankel function
$H_{\nu}^{(1)}=J_{\nu}+iY_{\nu}$. The Green function for
$\zeta<\zeta^{\prime}$ is \begin{equation} {\cal G}
(\zeta,\zeta^{\prime})=w^{-1}(\psi_{1}(\zeta)\psi_{2}(\zeta^{\prime})-\psi_{2}(\zeta)\psi_{1}(\zeta^{\prime}))
\end{equation}where the constant $w$
is the wronskian. The solution (33) satisfies the boundary
condition at $\zeta=\infty$. Namely, $\Psi(\infty)=0$ and
$\partial_{\zeta}\Psi(\zeta)(\infty)=0$ because ${\cal
G}(\zeta,\zeta)=0$ (these boundary conditions are imposed at
$\frac{k}{a}=\infty $ according to the definition of $\tau$ in eq.(22)).
The classical system (9) is non-Hamiltonian. Hence, strictly
speaking cannot be quantized by means of the  standard methods.
Its proper way of quantization is by means of the Lindblad theory
of quantum dissipative systems. However, for a small dissipation
$\gamma$ and small $H$ the wave equation (9) can be transformed
into a wave equation of a harmonic oscillator with a time
dependent frequency. The quantization of this oscillator (as in
refs.\cite{mukhanov}\cite{sakai}\cite{basset}) determines the
quadratic fluctuations
\begin{equation}
\langle \delta\phi_{q}^{2}\rangle \simeq\tau^{2\mu}\vert
H_{\nu}^{(1)}(\zeta)\vert^{2}\simeq k^{-2\nu}
\end{equation}
(for a small $k$). If $\Gamma$ is small then from eq. (32)
\begin{equation}
\nu=\frac{3}{2}+3\epsilon-\eta+\frac{3}{2}\Gamma(1-\sigma)-\epsilon
\Gamma(\frac{3}{2}\sigma-1)+\frac{3}{4}\Gamma^{2}
\end{equation}
At $\Gamma=0$ the formula (39) coincides  with the well-known
result. The surprising consequence of eqs.(38)-(39) is that at a
linear approximation to the indices of the quantum power spectrum
the effect of friction and warm inflation (noise) disappears in
the model (10) of refs.\cite{berera}\cite{adv} with $\sigma=1$ (
for this reason we included higher order terms in eq.(39)).

 The
thermal fluctuations are calculated from eqs.(29) and(33)
\begin{equation}\begin{array}{l}
\langle \delta\phi_{th}^{2}\rangle
=\gamma^{2}\beta^{-1}k^{-3}\int_{\zeta}^{\infty}
(\frac{\zeta}{\zeta^{\prime}})^{2\alpha}\Big({\cal
G}(\zeta,\zeta^{\prime})\Big)^{2}d\zeta^{\prime}
\end{array}\end{equation} For small $k$ we can set the lower limit in the
integral (40) to zero. Then (as $Y_{\nu}(\zeta)\simeq
\zeta^{-\nu}$ for a small $\zeta$)
\begin{equation}\begin{array}{l}
\langle \delta\phi_{th}^{2}\rangle
=\gamma^{2}\beta^{-1}k^{-3}\int_{0}^{\infty}
(\frac{\zeta}{\zeta^{\prime}})^{2\alpha}\Big({\cal
G}(\zeta,\zeta^{\prime})\Big)^{2}d\zeta^{\prime} \cr\simeq
k^{-3}\zeta^{2\mu}Y_{\nu}^{2}(\zeta)\simeq k^{-3}\zeta^{2\mu-2\nu}
\end{array}\end{equation}
For a small $\Gamma$ we have from eqs.(30) and (32)
\begin{equation}
\langle \delta\phi_{th}^{2}\rangle\simeq k^{-3-4\epsilon+2\eta
+3\sigma\Gamma+\epsilon\Gamma(3\sigma+1)-\frac{3}{2}\Gamma^{2}}
\end{equation}
Hence, surprisingly  in a linear approximation the index of the
thermal power spectrum  does not depend on $\Gamma$ if the
mysterious $\frac{3}{2}\gamma^{2}H\phi$  term is absent (
$\sigma=0$) in eq.(10).
\section{Summary}
We have calculated fluctuations in an Einstein-Klein-Gordon system
interacting with an environment (described by a Gaussian noise).
The resulting stochastic equations take a form depending on the
special environment of scalar fields interacting linearly with the
inflaton (in particular this leads to the
$\frac{3}{2}\gamma^{2}H\phi$ term in eq.(10)). The fluctuations
arise from  quantization as well as from the interaction with the
environment (thermal fluctuations). In an inflationary expansion
both fluctuations are nearly scale invariant. The spectral indices
in eqs.(38) and (41) are close to each other. The amplitude of
thermal fluctuations depends on the friction $\gamma$ (there are
some estimates on $\gamma$ in the warm inflation models
\cite{warm}). If this amplitude is of the same order as the
quantum one then it may be difficult to distinguish on the basis
of CMB measurements quantum and thermal fluctuations.
 The procedure which we have applied to calculate the inflaton fluctuations
follows the  standard  one \cite{basset}\cite{marozzi}\cite{hwang}
using an expression of metric fluctuations in terms of the  scalar
fields fluctuations. Our results for the spectral index are
different from earlier results of ref.\cite{moss}. The reason may
be that the authors \cite{moss} do not base their methods solely
on the Einstein-Klein-Gordon equations but on
 some thermodynamic arguments which may
involve another form of the energy-momentum tensor (the noise on
the rhs of the wave equation enforces a modification of the
energy-momentum tensor in order to satisfy the conservation law).
The impact of the $\frac{3}{2}\gamma^{2}H\phi$ term on the indices
of quantum and thermal fluctuations shows that the form of the
interaction of the inflaton with the environment has an essential
influence on these indices. The result could be tested in the CMB
observations as one does with the standard indices $\epsilon$ and
$\eta$.

{\bf Acknowledgement}

The author thanks the anonymous referee for pointing out an error in ref.\cite{adv}.

\end{document}